\documentclass[12pt]{iopart}

\usepackage{iopams}
\usepackage{graphicx}

\begin{document}
\title{Superconductivity of lanthanum revisited: enhanced critical temperature in the clean limit}

\author{P.~L\"{o}ptien}
\address{Institute of Applied Physics, Universit\"{a}t Hamburg, D-20355 Hamburg, Germany}
\author{L.~Zhou}
\ead{lzhou@physnet.uni-hamburg.de}
\address{Institute of Applied Physics, Universit\"{a}t Hamburg, D-20355 Hamburg, Germany}
\author{A.~A.~Khajetoorians}
\ead{akhajeto@physnet.uni-hamburg.de}
\address{Institute of Applied Physics, Universit\"{a}t Hamburg, D-20355 Hamburg, Germany}
\author{J.~Wiebe}
\ead{jwiebe@physnet.uni-hamburg.de}
\address{Institute of Applied Physics, Universit\"{a}t Hamburg, D-20355 Hamburg, Germany}
\author{R.~Wiesendanger}
\address{Institute of Applied Physics, Universit\"{a}t Hamburg, D-20355 Hamburg, Germany}

\begin{abstract}
The thickness dependence of the superconducting energy gap $\Delta_{\rm{La}}$ of double hexagonally close packed (dhcp) lanthanum islands grown on W(110) is studied by scanning tunneling spectroscopy, from the bulk to the thin film limit. Superconductivity is suppressed by the boundary conditions for the superconducting wavefunction at the surface and W/La interface, leading to a linear decrease of the critical temperature $T_c$ as a function of the inverse film thickness. For thick, bulk-like films, $\Delta_{\rm{La}}$ and $T_c$ are 40\% larger as compared to literature values of dhcp La measured by other techniques. This finding is reconciled by examining the effects of surface contamination as probed by modifications of the surface state, suggesting that the large $T_c$ originates in the superior purity of the samples investigated here.
\end{abstract}

%
\vspace{2pc}
\maketitle
%

\section{Introduction}
Commonly, the energy gap $\Delta$ of a superconducting film is determined in tunneling experiments where a conducting electrode and the superconductor are separated by an insulating layer. In early tunneling experiments, planar tunneling junctions with a thick oxide layer \cite{Shen1972, Lou1972, Wuehl1973} or point contacts~\cite{Levinstein1967} were utilized to determine the superconducting properties of many typical superconductors. With the advancement of low temperature scanning tunneling spectroscopy (STS) in ultra-high vacuum (UHV), it has become possible to probe \textit{in-situ} fabricated superconductors, and to determine $\Delta$ with atomic-scale spatial resolution \cite{deLozanna1985}. Here, $\Delta$ can be determined with a high degree of accuracy as a result of higher sample quality resulting from \textit{in-situ} preparation, and utilizing the vacuum barrier which acts as a perfect insulator. In recent years, such investigations have shown that seemingly well understood, elemental BCS superconductors, like Pb or In, behave dramatically different at the thin film limit.  For example, the onset of quantum well states allows superconductivity to remain robust in the thin film limit \cite{Eom2006, Zhang2010}, or even persist down to a single layer \cite{Zhang2010}.  For other systems, like Pb-Bi alloys, it has been shown that $T_c$ can be engineered by modifying the Fermi wavevector $k_{\rm{F}}$ \cite{Ozer2007}.  A thorough understanding of superconductivity at these length scales is not only interesting from an academic point of view, but also required for possible applications of superconductivity in nanoscale devices, mandating local probe investigations with high spatial and energy resolution.

Although the preparation quality is especially crucial for reactive materials as, e.g., lanthanides, this technique has not been applied to lanthanum yet, one of the few elemental superconductors with $T_c > 4$~K \cite{Buzea2005}.
The superconductivity of bulk La was investigated in a number of experimental studies \cite{Levinstein1967, Johnson1967, Maple1969, Shen1972, Lou1972, Wuehl1973, Pan1980}, including planar tunneling \cite{Lou1972, Shen1972, Wuehl1973} and point contact spectroscopy \cite{Levinstein1967}, and by theory \cite{Nixon2008, Bagci2010}. Most notably, La is an intermediate-coupling superconductor \cite{Levinstein1967, Johnson1967, Lou1972, Shen1972, Wuehl1973, Pan1980, Bagci2010} with a critical temperature at atmospheric pressure of $T_c^\mathrm{lit} = (4.98 \pm  0.04)$~K and $(6.04 \pm 0.07)$~K for the stable double hexagonally close packed (dhcp) and metastable fcc phase, respectively, and an extraordinary enhancement under compression \cite{Maple1969, Wuehl1973}. Studies about the low-dimensional properties of La are lacking.

In this letter, we report on a STS study of the superconducting properties of dhcp La islands. We observe that $\Delta_\mathrm{La}$ and $T_c$ of La are larger than commonly believed for clean bulk La~\cite{Levinstein1967, Johnson1967, Lou1972, Shen1972, Wuehl1973, Pan1980, Bagci2010}. Approaching the thin film limit, namely where the film thickness is comparable to the coherence length $\xi_0$, we determine a monotonous decrease of the superconducting properties, in agreement with a theoretical model that considers the boundary condition for the superconducting wavefunction \cite{Simonin1986}, ruling out significant quantum size effects in the superconductivity.  We investigated samples of different purity and show that a reduction in $T_c$ is correlated with increased surface contamination and quenching of the unoccupied surface state of the La islands.

\section{Experimental Procedures}
\textit{In-situ} prepared La films were studied in a commercially available UHV STM \cite{SPECS} at a base temperature of $T = 1.2$~K, unless otherwise specified, and a base pressure $p < 2\times 10^{-10}$~mbar. The W(110) surface was cleaned by cycles of annealing in an oxygen atmosphere and subsequent thermal flashing \cite{Bode2007}. La films were grown by electron beam evaporation on a clean W(110) substrate held at room temperature. Afterwards, each sample was annealed for 5 to 12 minutes at temperatures in the range of 700 to $800\,^\circ\mathrm C$ and slowly cooled to room temperature, avoiding a rapid thermal quenching to bypass the metastable fcc phase. 
Two types of La sources with different nominal purity were used for the experiments \cite{MateckChempur}. The corresponding La films are named 1$^\mathrm{st}$ and 2$^\mathrm{nd}$ generation samples in the following.
STM topographs were recorded in constant-current mode, with a sample bias voltage $V = 1$~V and a tunneling current in the range of $I = 30$ to $200$~pA. STS was performed using a standard lock-in technique, adding a modulation voltage $V_\mathrm{mod}$ to $V$.
The $\mathrm dI/\mathrm dV$ spectra dedicated to study superconductivity were taken using normal metal and superconducting Nb tips \cite{Wiebe2004} with $I_\mathrm{stab} = 100$ to 150~pA at $V_\mathrm{stab} = -6$~mV, $V_\mathrm{mod} = 0.04$ to 0.07 mV. In this junction resistance range, we did not observe any effects of Josephson supercurrents or Andreev reflections. The spectra for the surface state were recorded with $I_\mathrm{stab} = 500$~pA at $V_\mathrm{stab} = +1$~V, $V_\mathrm{mod} = 1$~mV.

\begin{figure}[htbp]
	\centering
		\includegraphics[width=0.8\textwidth]{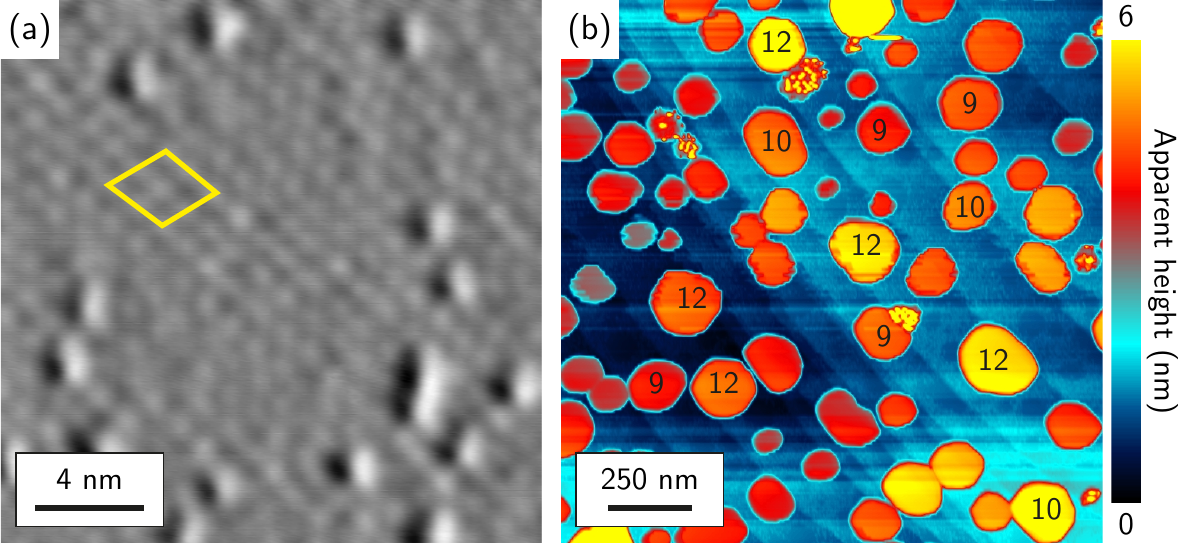}
	\caption{STM topographs of La/W(110).
	(a)~Laterally differentiated topograph of the wetting layer in between the La islands with the rhombic unit cell indicated in yellow. 
	(b)~Islands of a $1^\mathrm{st}$ generation sample with heights in ML as depicted by the numbers.}
	\label{fig:Figure_1}
\end{figure}

Annealing of La/W(110) leads to Stranski-Krastanov growth~\cite{Wegner_all}, i.e., a La wetting layer (WL) with a thickness of one monolayer (ML) on the W(110) surface [Fig.~\ref{fig:Figure_1}(a)],
while the additional material forms flat-top dhcp La(0001) islands [Fig.~\ref{fig:Figure_1}(b)]. Height and lateral extension of the islands depend on the combination of deposition time and rate, and annealing time and temperature. In total, islands with thicknesses $d$ in the range between $d = 2.5$~nm and $d =140$~nm (8 to 460~ML) were grown, which covers a broad range from the thin film to the bulk limit, with respect to the coherence length in the clean limit, \mbox{$\xi_0^\mathrm{lit} = 36.3$~nm} \cite{Pan1980}. In order to avoid lateral size effects, we only studied islands with a diameter $\gg \xi_0^{\mathrm{lit}}$.

\section{Determination of the energy gap}

\begin{figure}[htbp]
	\centering
		\includegraphics[width=0.8\textwidth]{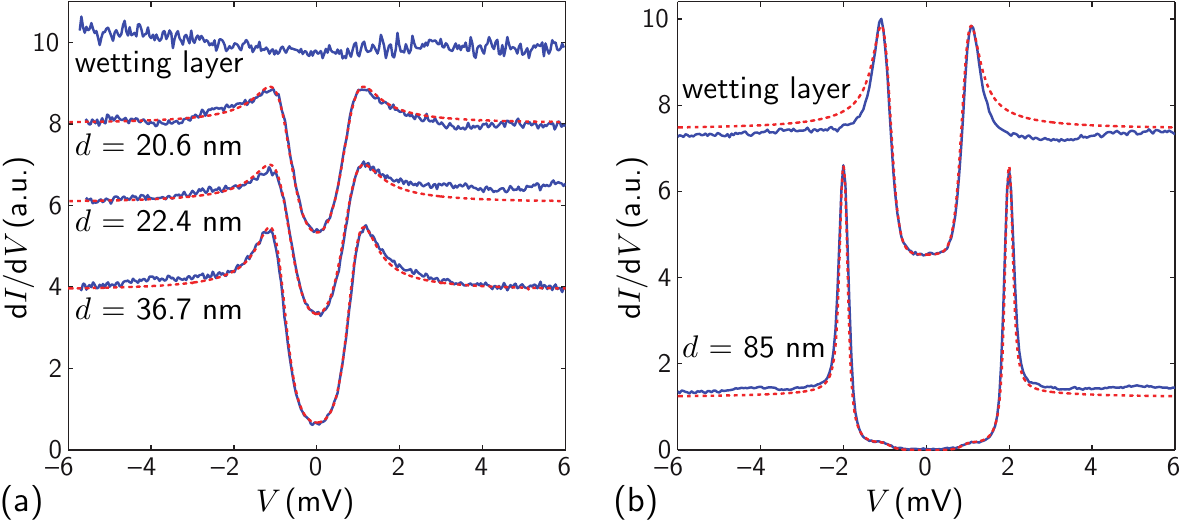}
	\caption{Determination of $\Delta_\mathrm{La}(T,d)$ for La islands with different thicknesses $d$, using (a) normal metal and (b) superconducting Nb tips. Experimental curves (blue solid lines) were taken on the WL to characterize the tip DOS and with the same tip on the La islands with indicated $d$. The fitted calculations are shown as red dotted curves. (a)~Experimental data: \mbox{$V_\mathrm{mod} = 0.06$~mV}, $T = 1.23$~K.	Fitted calculations: \mbox{$V_\mathrm{mod,eff} = 0.10$~mV}, \mbox{$\Delta_\mathrm{La} = 0.79$}, $0.80$, $0.89$~meV, $\Gamma_\mathrm{La} = 0.25$, $0.25$, $0.14$~meV (from top to bottom).	(b)~Experimental data: $V_\mathrm{mod} = 0.07$~mV, $T = 1.14$~K. Fitted calculations: $V_\mathrm{mod,eff} = V_\mathrm{mod}$, $\Delta_\mathrm{tip} = 0.94$~meV, $\Gamma_\mathrm{tip} = 0.01$~meV, $\Delta_\mathrm{La} = 1.05$~meV, $\Gamma_\mathrm{La} = 0.07$~meV. The curves are vertically shifted for visual clarity.}
	\label{fig:Figure_2}
\end{figure}

STS on La islands reveals a symmetric superconducting gap, $\Delta_{\rm{La}}$, around $E_{\rm{F}}$ resulting from the superconductivity of the probed island [Fig.~\ref{fig:Figure_2}(a)], while the WL shows no gap, indicating a normal metal. Moreover, $\Delta_{\rm{La}}$ is reduced with decreasing $d$. Spectra taken with Nb tips [Fig.~\ref{fig:Figure_2}(b)] correspondingly show a gap on the WL stemming from the tip density of states (DOS), and a larger gap on the La islands stemming from the interplay of the tip and sample DOS.

In order to precisely extract $\Delta_{\rm{La}}$ for a given $d$, the experimental curves were each fitted with a numerically calculated differential conductance (Appendix A), involving a BCS-like DOS for the sample, or for both electrodes in the case of Nb tips,
\begin{equation}
	N_\mathrm{sc}(E, \Gamma) = N_n\ \Re\left(\frac{E - i\,\Gamma}{\sqrt{(E - i\,\Gamma)^2-\Delta^2}}\right)\ .
	\label{BCS_DOS_1_gamma}
\end{equation}

Here, $N_n$ is the DOS of the electrode in its normal metal state which is assumed to be constant, $E$ is the energy, $\Delta$ is the familiar energy gap from BCS theory, and $\Gamma$ is a broadening parameter which was originally introduced to describe the finite lifetimes of quasiparticles in the tunneling process \cite{Dynes1978}. The fitted calculations (Appendix A) yield $\Delta_\mathrm{La}$ and $\Gamma_\mathrm{La}$ as free fit parameters, while $\Delta_\mathrm{Nb}$ and $\Gamma_\mathrm{Nb}$ were determined from the WL spectra taken with the same micro-tip. The excellent fit quality [Fig.~\ref{fig:Figure_2}(a,b)] permits an accurate determination of these quantities, resulting in values of up to $\Delta_{\rm{La}} = 1.05$~meV and $\Gamma_{\rm{La}} \approx 0.1$ to $0.6$~meV. There is no strong lateral variation of superconductivity on the islands, at least for distances $\geq \xi_0^\mathrm{lit}$ from the islands' edges (Appendix B). Hence, we conclude that $\Delta_{\rm{La}}$ depends only on $d$ and $T$.  The obtained $\Delta_\mathrm{La}(T,d)$ is shown in Fig.~\ref{fig:Figure_3}(a) for both sample generations. Below, we first focus on the results from the cleaner $1^\mathrm{st}$ generation samples, and discuss the effect of impurities on superconductivity later. Going from the bulk limit $d\gg\xi_0^\mathrm{lit}$ to the thin film limit, the values of $\Delta_\mathrm{La}(T,d)$ show a linear decrease as a function of inverse thickness with no obvious saturation above $d=\xi_0^\mathrm{lit}$.

\section{Determination of the critical temperature}

$T_c(d)$ is determined from the experimental $\Delta_\mathrm{La}(T,d)$ using BCS theory (Appendix C). In the La bulk limit, the relation between the zero-temperature energy gap $\Delta_\mathrm{La}(0,d)$ and $T_c(d)$ was shown to equal $2\Delta_\mathrm{La}(0,d)/k_\mathrm BT_c(d)=3.75 \pm 0.02$~\cite{Lou1972, Shen1972, Wuehl1973, Pan1980}. We assume the same constant value for the whole range of La thicknesses studied here. In fact, the only two effects which could alter its value for decreasing $d$ are (i) phonon softening \cite{Dickey1968}, which would emerge as a shift of phonon modes in experimental $\mathrm dI/\mathrm dV$-curves~\cite{Brun2009}, and (ii) electronic quantum size effects, which would appear as discrete states in the higher voltage range of the spectra. There are no clear indications for such features (Fig.~\ref{fig:Figure_2}, Fig.~\ref{fig:Figure_4}). Therefore, both effects can be ruled out, and should only emerge in very thin films of up to a couple of monolayers (5 to 10~ML in Pb \cite{Brun2009}). For our thinnest films, where $T_c$ becomes comparable to $T$, the finite experimental temperature is considered by calculating $\Delta_\mathrm{La}(0,d)$ from $\Delta_\mathrm{La}(T,d)$ by a numerical integration according to BCS theory (Appendix C).

\begin{figure}[htbp]
	\centering
		\includegraphics[width=0.80\textwidth]{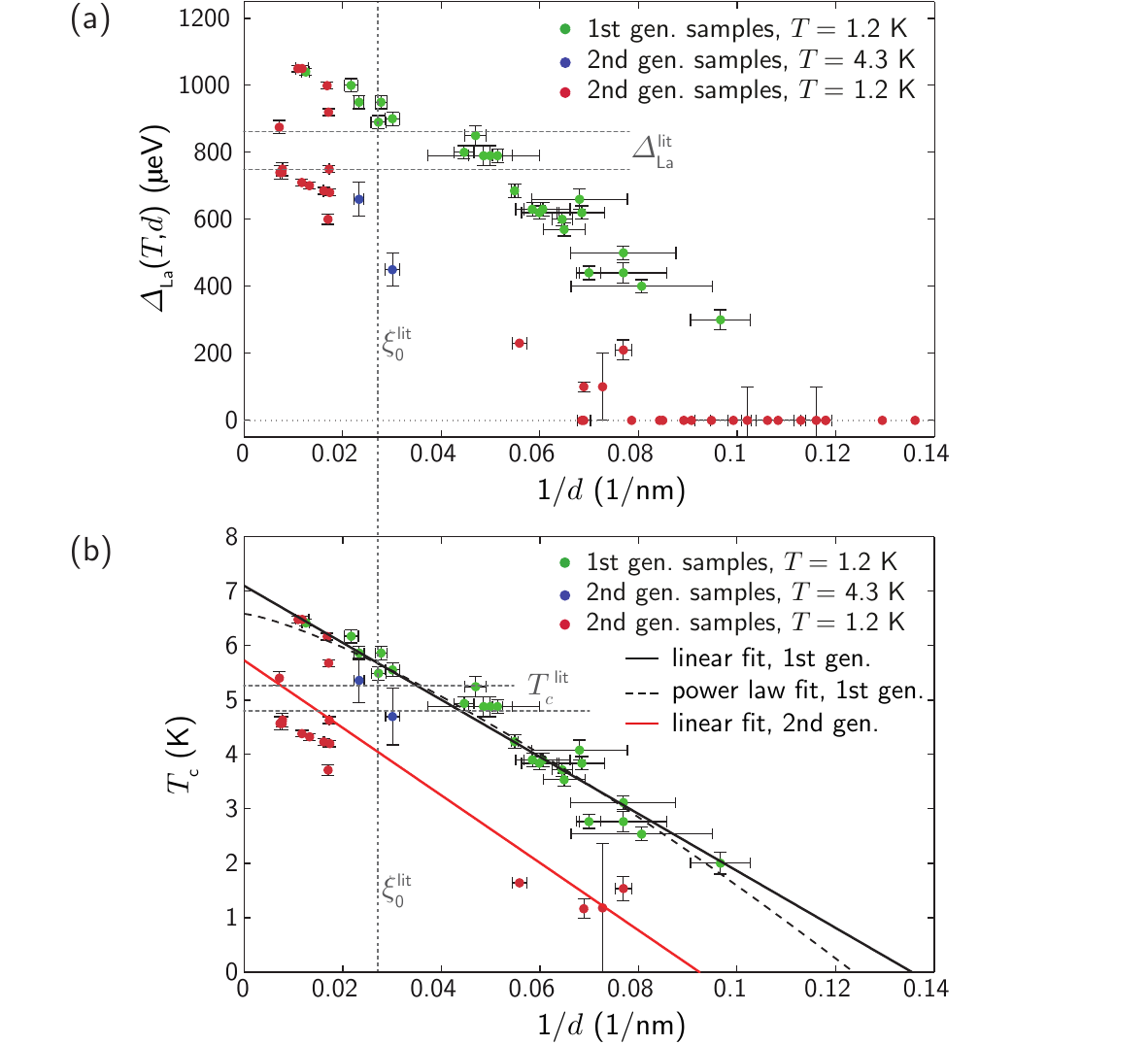}
		\caption{Dependence of (a) $\Delta_\mathrm{La}(T,d)$ and (b) $T_c(d)$ on the inverse film thickness $1/d$. Sample generations and measurement temperatures are indicated. The error bars are due to uncertainties in the measured film thickness and in the fit parameters. The literature bulk values $\xi_0^\mathrm{lit}$, $\Delta_{\rm{La}}^\mathrm{lit}$, and $T_c^\mathrm{lit}$ are indicated by vertical and horizontal lines, respectively, in (a,b).	Dashed and solid curves in (b) show power law fits and linear fits according to the Simonin model (Eq.~\ref{eq:T_c_vs_d}), respectively.}
	\label{fig:Figure_3}
\end{figure}

The obtained $T_c(d)$ shows a linear behavior in $1/d$ with highest values of $T_c=6.5$~K for the thickest islands [Fig.~\ref{fig:Figure_3}(b)]. Note that $T_c(d)$ calculated from the $4.3$~K experimental data roughly coincides with $T_c(d)$ calculated from the $1.2$~K data, justifying the determination of $T_c(d)$ via BCS theory.

\section{Thickness dependence and comparison to bulk values}
In order to explain the observed thickness dependence of $T_c$, at first the very general relation $T_c(d) \propto (1/d)^\alpha$ is fitted to the experimental data [dashed line in Fig.~\ref{fig:Figure_3}(b)]. The resulting fit parameter, $\alpha = 1.29 \pm 0.46$, is close to $\alpha = 1$. This strongly suggests that the experimental data can be explained by the so-called Simonin model~\cite{Simonin1986}, where the boundary conditions for the superconducting wavefunction, imposed by the interface to the W substrate and the surface, corresponds to an additional term in the Ginzburg-Landau free energy. This leads to a reduction in $T_c$ for thin superconducting films, resulting in
\begin{equation}
	T_c(d) = T_{c,\mathrm{bulk}} \left(1 - \frac{d_c}{d}\right)\ .
	\label{eq:T_c_vs_d}
\end{equation}

Here, the critical thickness $d_c$ can be interpreted as a threshold thickness, which is required for a La film to develop superconductivity, if the Simonin model holds for such very thin films. Keeping $T_{c,\mathrm{bulk}}$ and $d_c$ as free parameters, Eq.~\ref{eq:T_c_vs_d} is fitted to the experimental data [black solid line in Fig.~\ref{fig:Figure_3}(b)]. The fitted value for the critical thickness is \mbox{$d_c = (7.38 \pm 1.04)$~nm} (about 25~ML in the [0001] direction). Note that, although no superconducting gap is observed for $d \leq 10$~nm under our experimental conditions, it cannot be ruled out that these islands have a finite $T_c$ which is below the measurement temperature of 1.2~K. An expected value for $d_c$ can be determined via $d_{c}^\mathrm{lit} = 2/[k_\mathrm{TF} N(0)V]$ \cite{Simonin1986} from literature values for the inverse Thomas-Fermi screening length $k_{\rm{TF}} = (1.543 \pm 0.036)$~nm$^{-1}$, and for the product of the electronic density of states at the Fermi level and the electron-phonon coupling potential $N(0)V = (0.286 \pm 0.006)$ \cite{Johnson1967, Pan1980, Bagci2010}, giving $d_{c}^\mathrm{lit} = (4.54 \pm 0.09)\,\mathrm{nm}$, which is close to the experimental result. The relatively large value of $d_{c}$, compared to other superconducting materials such as Pb, thus originates from the comparatively small Fermi velocity~\cite{Pan1980} of La.

The value for the bulk critical temperature extracted from the fit to the Simonin model is \mbox{$T_{c,\mathrm{bulk}} = (7.10 \pm 0.38)$~K}. The aforementioned relation $2\Delta_\mathrm{La}(0,d)/k_\mathrm BT_c(d)$ then yields a bulk energy gap of $\Delta_\mathrm{La,bulk}(0) = (1.15 \pm 0.06)$~meV, which coincides with the value extrapolated from Fig.~\ref{fig:Figure_3}(a). As a main result of this work, both $T_{c,\mathrm{bulk}}$ and $\Delta_\mathrm{La,bulk}(0)$ are almost 40\% higher in comparison to reported bulk values \cite{Levinstein1967, Johnson1967, Lou1972, Shen1972, Wuehl1973, Pan1980, Nixon2008, Bagci2010}. This result is surprising, since in previous experimental studies of the thickness-dependent $T_c(d)$ in other materials, which also revealed a linear dependence in agreement with the Simonin model, the extrapolated values usually are in agreement with the respective bulk values \cite{Ozer2007, Fabrega2011a, Song2011a}.

In the following, we discuss the origin of the large $T_c$ value. The effect of strain on the La islands can be neglected in the thickness dependency $T_c(d)$, even though La is a relatively soft metal: For Gd/W(110), which has a similar lattice constant to La/W(110), only the monolayer and bilayer have slightly enhanced lattice constants of 2\% and 0.3\% \cite{Nepijko2000}, while thicker films are already relaxed to the bulk lattice parameter. Moreover, the residual strain is tensile in nature and would lead to a reduction in $T_c$ \cite{Lou1972, Wuehl1973}. Therefore, we can rule out that the experimentally observed increase in $\Delta_\mathrm{La,bulk}(0)$ and $T_{c,\mathrm{bulk}}$ is related to strain effects.

\section{Effect of purity on $\Delta$ and $T_{c}$}

\begin{figure}[htbp]
	\centering
		\includegraphics[width=0.80\textwidth]{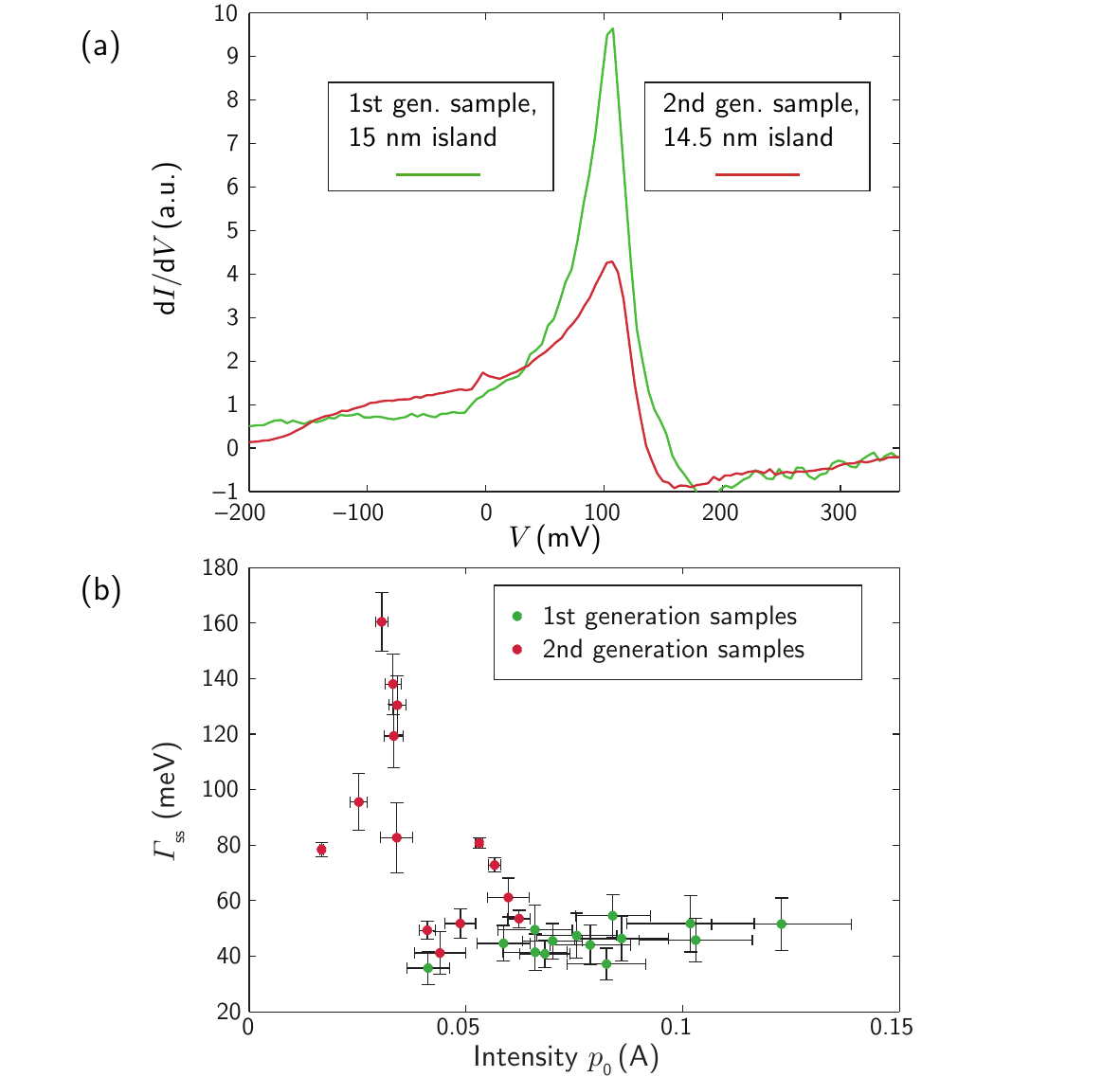}
	\caption{(a) STS spectra indicating the $d_{z^2}$-like surface state that forms on La(0001). The curves are normalized such that the differential conductance at \mbox{$V_\mathrm{stab} = +1$~V} coincides. The negative $\mathrm dI/\mathrm dV$ values originate from an interplay between the strongly peaked sample DOS and a tip DOS with negative slope \cite{Wegner_all}. (b)~Quantitative analysis of FWHM $\Gamma_\mathrm{ss}$ and intensity $p_0$ of the surface state extracted from Lorentzian fits to STS spectra like shown in (a) (Appendix D).}
	\label{fig:Figure_4}
\end{figure}

Since impurities are known to play a crucial role for superconductivity, we finally investigate the effect of purity of the sample on $\Delta_\mathrm{La}(T,d)$ and $T_c(d)$ by an investigation of the $2^\mathrm{nd}$ generation samples, where STM topographs revealed residual surface contamination. The surface contamination is quantified by the intensity and width of the $d_{z^2}$-like surface state which forms on the lanthanide (0001) surfaces \cite{Getzlaff1999, Wegner_all} and is very sensitive to adsorbates \cite{Getzlaff1999, Wegner_all}. The spectra in Fig.~\ref{fig:Figure_4}(a) indicate a well developed surface state by a strong resonance at $V \approx 0.1$~V in the $1^\mathrm{st}$ generation samples, while the surface state is partly quenched in the $2^\mathrm{nd}$ generation samples. This finding is supported by a quantitative analysis (Appendix D) of the full width at half maximum (FWHM) $\Gamma_\mathrm{ss}$ and intensities $p_0$ of the resonances measured on all investigated samples [Fig.~\ref{fig:Figure_4}(b)]. For the $1^\mathrm{st}$ generation samples, $\Gamma_\mathrm{ss}$ has a value comparable to or even lower than the reported intrinsic lifetime broadening due to electron-electron and electron-phonon scattering for very clean samples \cite{Wegner_all} (Appendix D). In contrast, on the $2^\mathrm{nd}$ generation samples, the surface state lifetime is further reduced by defect scattering as revealed by a roughly doubled $\Gamma_\mathrm{ss}$. We can therefore conclude that the surface purity of the $2^\mathrm{nd}$ generation samples is strongly reduced with respect to that of the $1^\mathrm{st}$ generation samples.

As shown in Fig.~\ref{fig:Figure_3}(a,b), the reduced purity of the $2^\mathrm{nd}$ generation samples is correlated with overall reduced values of $\Delta_\mathrm{La}(T,d)$ and $T_c(d)$. Eq.~\ref{eq:T_c_vs_d} fitted to this experimental data results in \mbox{$T_{c,\mathrm{bulk}}^* = (5.73 \pm 0.62)$~K} and \mbox{$d_c^* = (10.82 \pm 3.36)$~nm} [red straight line in Fig.~\ref{fig:Figure_3}(b)]. Since $d_c^*$ enters the boundary condition for the suppression of the superconducting order parameter \cite{Simonin1986}, the enhanced value of $d_c^*$ indicates a stronger decay of the superconducting order parameter at the dirtier surface boundary of the $2^\mathrm{nd}$ generation samples. These results suggest that the superior sample quality of the $1^\mathrm{st}$ generation samples with respect to previous studies \cite{Shen1972, Lou1972, Wuehl1973, Levinstein1967, Johnson1967, Pan1980, Maple1969, Bagci2010, Nixon2008} is responsible for the enhanced values of $\Delta_\mathrm{La,bulk}(0)$ and $T_{c,\mathrm{bulk}}$ of these samples. This result is astonishing: While the surface state and the observed contamination are localized in the topmost atomic layer of the La islands, and thus, their superconductivity is expected to be influenced by the surface only within a region of $\mbox{$\xi_0^\mathrm{lit} = 36.3$~nm}$, we observe a reduction of $\Delta_\mathrm{La,bulk}(0)$ and $T_{c,\mathrm{bulk}}$ even for the thickest islands with $d=140$~nm. 
However, it is likely that not only the surface, but also the interior and the W-La interface of the $2^\mathrm{nd}$ generation La islands are dirtier than for the $1^\mathrm{st}$ generation samples. This might explain the correlation between $T_c$ and surface contamination for the thick islands with $d>\xi_{0}^\mathrm{lit}$, where the surface is not expected to play a crucial role for the superconducting properties.

\section{Summary}
In summary, our observations reveal that the intrinsic bulk energy gap and critical temperature of dhcp La are 40\% larger as compared to the values cited in the literature. In addition, we quantitatively determined the thickness dependence of $\Delta_{\rm{La}}$ and $T_c$, which is in good agreement with a theoretical model that considers the boundary conditions for the superconducting wavefunction.  We find that superconductivity does not persist below a critical thickness of 25~ML. We consider the effects of sample purity, as correlated with modifications to the unoccupied surface state of La, and find, regardless of the thickness, that superconductivity is reduced at increased surface contamination. Our results suggest, that a superior purity of the samples investigated here explains why we observe an enhancement of $T_c$ as compared to previous reports. This highlights the challenge in the investigation of the superconducting properties of the notoriously reactive lanthanides.

\section*{Acknowledgments}
We acknowledge financial support from the ERC Advanced Grant ``ASTONISH'', and from the DFG via Graduiertenkolleg 1286. A.A.K. acknowledges Project No. KH324/1-1 from the Emmy-Noether-Program of the DFG. We thank A. Kamlapure for fruitful discussions.

\section*{Appendix A: Numerical calculation of the differential conductance}
The energy gap of lanthanum was investigated by STS with superconducting Nb tips and normal metal tungsten and PtIr tips. A lock-in technique gives direct access to $\mathrm dI/\mathrm dV$. There, the finite modulation voltage $V_\mathrm{mod}$ (RMS value) limits the energy resolution and broadens the coherence peaks, which adds to the usual thermal broadening. In order to determine the superconducting energy gap $\Delta$ and lifetime broadening parameter $\Gamma$, the differential conductance is calculated numerically in analogy to the working principle of the lock-in amplifier \cite{Wiebe2004}:
\begin{equation}
	\frac{\mathrm dI}{\mathrm dV}(V)\propto \int_{-\pi/2}^{+\pi/2} \sin(\alpha)\, I\left(V+\sqrt{2}\,V_\mathrm{mod}\,\sin(\alpha),T\right)\,\mathrm d\alpha\ .
	\label{dIdV_SC_2}
\end{equation}

In this formula, $I$ is the tunneling current, defined as
\begin{equation}
	I(V, T)\propto \int_{-\infty}^{+\infty} N_1(E)\,N_2(E+eV)\, \left[f(E + eV, T)-f(E, T)\right]\, \mathrm d E\ .
	\label{IV_SC}
\end{equation}

$N_1(E)$ and $N_2(E)$ are the densities of states (DOS) of the two tunneling electrodes, and $f(E, T)$ is the Fermi function.

In the simplest case, where the tunneling junction consists of a superconducting and a normal metal electrode (i.e., superconducting La island and normal metal tip \textsc{or} superconducting tip and normal metal wetting layer), the DOS of the superconducting electrode, defined by Eq.~1 of the main paper, is used as $N_1(E)$. Furthermore, we assume that the DOS $N_2(E)$ of the normal metal electrode is constant on the relevant scale of $\Delta_1$. In that case, a fitting of Eq.~\ref{dIdV_SC_2} to the experimental data, with only $\Delta_1$ and $\Gamma_1$ as free parameters, yields these quantities with good accuracy.

In STS measurements with two superconducting electrodes (i.e., superconducting La island and superconducting tip), $N_1(E)$ is the same as before. We assume a similar DOS for the second superconducting electrode, $N_2(E)$, with its normal metal DOS $N_{n,2}$. In addition, we consider the parameters $\Delta_2$, $\Gamma_2$, and the applied voltage:
\begin{equation}
	N_2(E+eV) = N_{n,2}\ \Re\left(\frac{E + eV + \sqrt 2\, eV_\mathrm{mod}\,\sin(\alpha)- i\,\Gamma_2}{\sqrt{\left[E + eV + \sqrt 2\, eV_\mathrm{mod}\,\sin(\alpha)- i\,\Gamma_2\right]^2-\Delta_2^2}}\right)
	\label{eq:rho_2}
\end{equation}

As $\Delta_1$ and $\Gamma_1$ are already known from the characterization experiment we performed for each tip on the wetting layer, fitting Eq.~\ref{dIdV_SC_2} to the experimental data yields $\Delta_2$ and $\Gamma_2$ with good accuracy.

While most STS studies of superconductors fit experimental  $\mathrm dI/\mathrm dV$-curves by using only Eq.~\ref{IV_SC}, the approach introduced by Eq.~\ref{dIdV_SC_2} naturally considers a finite modulation voltage. Moreover, a possible electronic noise added to the bias voltage can be described by this approach when using an increased effective modulation voltage $V_\mathrm{mod,eff}\geq V_\mathrm{mod}$. This enables to decouple several effects that broaden the $\mathrm dI/\mathrm dV$-curves, and to determine the ``intrinsic'' lifetime broadening parameter $\Gamma$ for tip and sample.

\begin{figure}[htbp]
	\centering
		\includegraphics[width=0.80\textwidth]{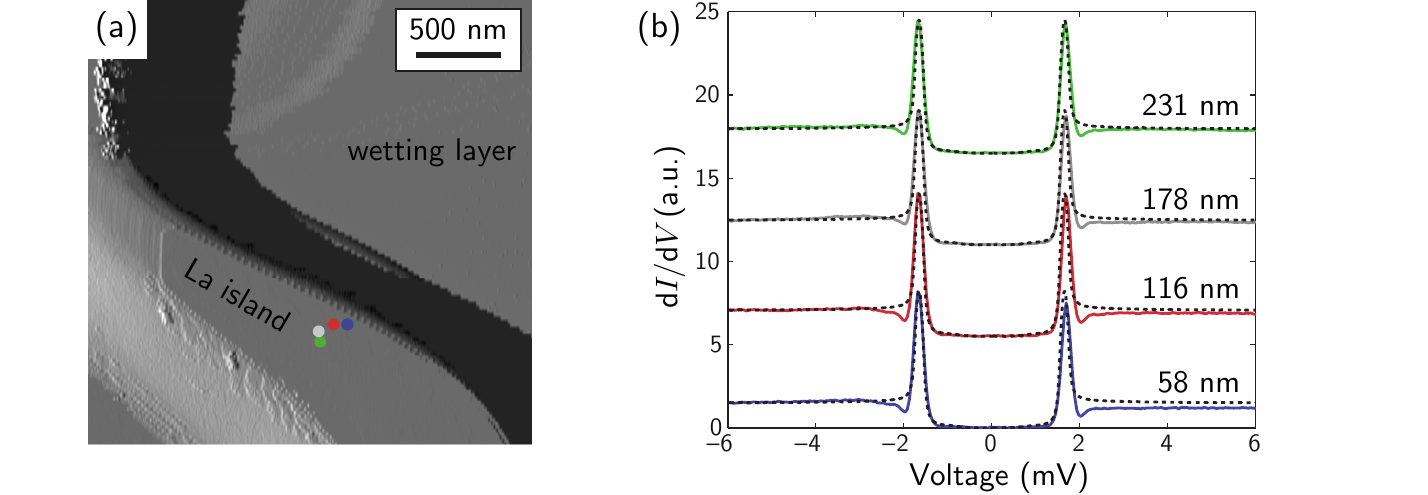}
	\caption{Investigation of a possible lateral variation of superconductivity. 
	(a)~STM topograph (laterally differentiated image, \mbox{$I = 50$~pA}, $V = 950$~mV, $T = 1.16$~K) of a 137~nm thick lanthanum island taken with a superconducting Nb tip ($\Delta_\mathrm{tip} = 0.770$~meV, \mbox{$\Gamma_\mathrm{tip} = 0.020$~meV}).
	(b)~$\mathrm dI/\mathrm dV$-curves taken at increasing separations from the rim, as depicted by the colored dots in (a) ($I_\mathrm{stab} = 100$~pA, $V_\mathrm{stab} = -6$~mV, $V_\mathrm{mod} = 0.080$~mV). The spectra are each shifted by 5.5~a.u. for better clarity.
Numerical calculations (dotted curves), obtained with $T = 1.16$~K and $V_\mathrm{mod,eff} = V_\mathrm{mod}$, reveal the very same parameters, $\Delta_\mathrm{La} = 0.875$~meV and $\Gamma_\mathrm{La} = 0.025$~meV, for all experimental $\mathrm dI/\mathrm dV$-curves. Slight deviations at positive voltages can be ascribed to an insufficient stabilization of the tunneling junction. The dips in $\mathrm dI/\mathrm dV$ at voltages above the coherence peaks originate from the Nb tip and are not related to the lanthanum island.}
	\label{fig:lateral_superconductivity}
\end{figure}

\section*{Appendix B: Exclusion of a lateral variation of the superconducting properties}
A possible lateral variation of the superconducting properties of lanthanum islands is inspected in a control experiment on a particular 137~nm thick island [Fig.~\ref{fig:lateral_superconductivity}(a,b)]. $\mathrm dI/\mathrm dV$-spectra were taken at different positions on the island [colored dots in Fig.~\ref{fig:lateral_superconductivity}(a)]. Even the blue point is 58~nm away from the rim, which is clearly above the coherence length. The $\mathrm dI/\mathrm dV$-spectra do not show any lateral variation on the relevant energy scale up to \mbox{$|eV| = 6$~meV} [Fig.~\ref{fig:lateral_superconductivity}(b)].
Hence, the superconducting properties of the lanthanum islands, $\Delta_\mathrm{La}$ and $\Gamma_\mathrm{La}$, are only a function of thickness $d$ and temperature $T$. 

\section*{Appendix C: Determination of the critical temperature}
\begin{figure}[htbp]
	\centering
		\includegraphics[width=0.80\textwidth]{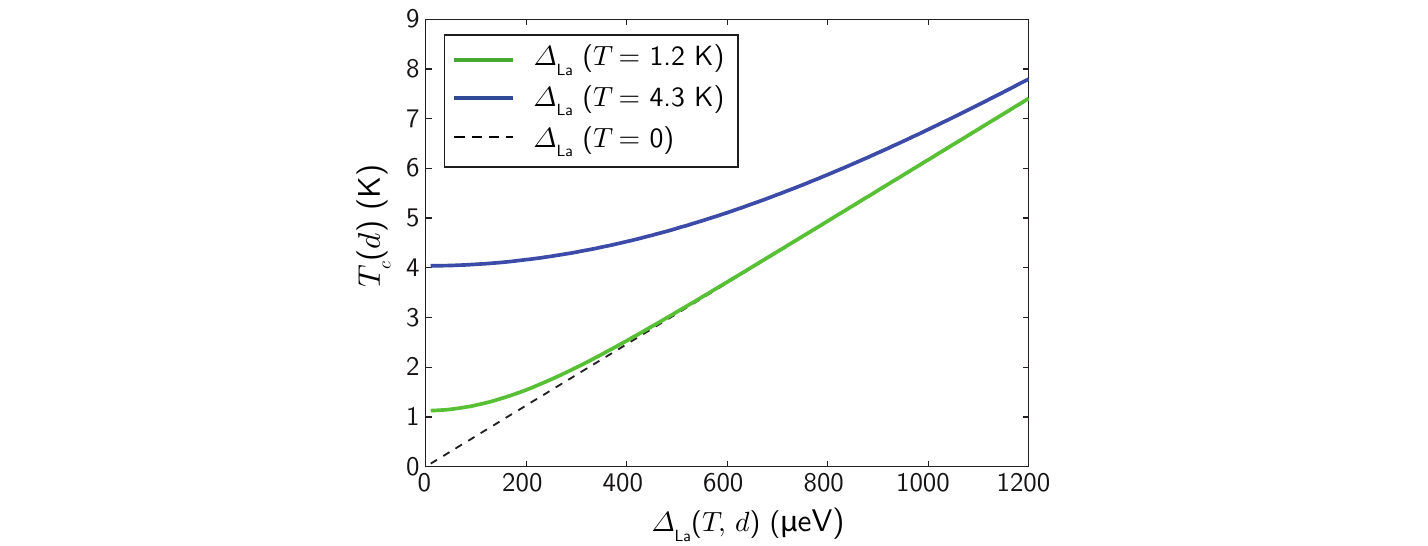}
	\caption{	Derivation of $T_c(d)$ from the experimental values of $\Delta_\mathrm{La}(T,d)$, for $T=0$, $T=1.2$~K, and $T=4.3$~K.}
	\label{fig:Figure_S_TcvsDelta}
\end{figure}

In the main paper, the following relation between $T_c(d)$ and $\Delta_\mathrm{La}(0, d)$ is given~\cite{Lou1972, Shen1972, Wuehl1973, Pan1980}:
\begin{equation}
	\frac{2\Delta_\mathrm{La}(0,d)}{k_\mathrm BT_c(d)} = 3.75 \pm 0.02
	\label{eq:fraction}
\end{equation}
In order to obtain $T_c(d)$ from the experimental $\Delta_\mathrm{La}(T, d)$, an understanding of the temperature dependence of $\Delta_\mathrm{La}(T, d)$ is required. For the intermediate-coupling superconductor lanthanum \cite{Johnson1967, Pan1980, Bagci2010}, the relations between $\Delta_\mathrm{La}(T, d)$ and $T_c(d)$ are given by 

\begin{equation}
	k_\mathrm B T_c(d) = 1.065\cdot\, \hbar \omega_\mathrm D(d)\cdot\, \mathrm e^{-1/[N(0)V]}
	\label{eq:Tc}
\end{equation}
\begin{equation}
	\frac{1}{N(0) V} = \int_0^{\hbar\omega_\mathrm D(d)} \frac{\tanh \frac{1}{2}\beta \sqrt{\xi^2 + \Delta_\mathrm{La}(T,d)^2}}{\sqrt{\xi^2 + \Delta_\mathrm{La}(T,d)^2}}\,\mathrm d\xi = \frac{1}{0.286 \pm 0.006}\ .
	\label{eq:Delta_T}
\end{equation}
Here, $\beta = 1/k_\mathrm BT$.
Eq.~\ref{eq:fraction} is used to determine $T_c(d)$ as a function of $\Delta_\mathrm{La}(0, d)$ [dashed line in Fig.~\ref{fig:Figure_S_TcvsDelta}]. The dependence of $T_c(d)$ as a function of $\Delta_\mathrm{La}(T, d)$ is calculated by a numerical integration of Eq.~\ref{eq:Delta_T} using Eq.~\ref{eq:Tc} in order to relate $\hbar\omega_\mathrm D(d)$ to $T_c(d)$.

In Fig.~\ref{fig:Figure_S_TcvsDelta}, $T_c(d)$ is shown as a function of $\Delta_\mathrm{La}(1.2\ \mathrm{K}, d)$ (green curve) and $\Delta_\mathrm{La}(4.3\ \mathrm{K}, d)$ (blue curve).
Concerning $T = 1.2$~K, in the range of the relevant data, $\Delta_\mathrm{La}(1.2\ \mathrm{K}, d) = 400\ \mathrm{to}\ 1000$~$\mu$eV, there is basically no deviation between $\Delta_\mathrm{La}(0, d)$ and $\Delta_\mathrm{La}(T, d)$, hence $T_c(d)$ is immediately given by $2\Delta_\mathrm{La}(T,d)/k_\mathrm BT_c(d) = 3.75 \pm 0.02$.
In contrast, $T = 4.3$~K leads to a significant deviation between $\Delta_\mathrm{La}(0,d)$ and $\Delta_\mathrm{La}(T,d)$.
The calculated relations are used to determine $T_c(d)$ for all measured $\Delta_\mathrm{La}(T,d)$, as shown in the main paper.

\section*{Appendix D: Contamination-induced quenching of the surface state}
As described in the main text, the $d_{z^2}$-like surface state which forms on La(0001) is sensitive to surface contamination (Fig.~4). In the following, the reduction of intensity and enhancement of broadening is quantitatively determined from the experimental STS curves of the two sample generations. This enables to relate these values to the surface state lifetime and quality of La films for comparison with previous experimental studies \cite{Wegner_all}. 
In order to describe the differential conductance $\mathrm dI/\mathrm dV(V)$ analytically, a simple model in analogy to \cite{Wegner_all} is introduced. 

Neglecting any voltage dependence of the transmission coefficient $\mathcal T(E, V)$, 
and assuming a constant electronic density of states of the tip, an expression for $\mathrm dI/\mathrm dV$ is given as
\begin{equation}
	\frac{\mathrm dI}{\mathrm dV}(V) \propto \int N_s(E)\, \mathcal T(E)\, f'(E-eV,T)\, \mathrm d E\ .
\end{equation}

Here, $N_s(E)$ is the density of states of the sample, $f'(E-eV,T)$ denotes the differentiation of the Fermi function with respect to $V$.
The transmission coefficient $\mathcal T(E)$ is rather constant.
The (negative) effective mass of the La(0001) surface state was determined by \textit{ab initio} calculations as $|m_\mathrm{eff}| > 2\,m_e$ \cite{Wegner_all}. 
Here, this rather weak dispersion is neglected completely, and $N_s(E)\,\mathcal T(E)$ is first approximated by a $\delta$-function $\delta(E-E_0)$, where $E_0$ equals the band maximum at $k=0$. 

\begin{figure}[htbp]
	\centering
		\includegraphics[width=0.80\textwidth]{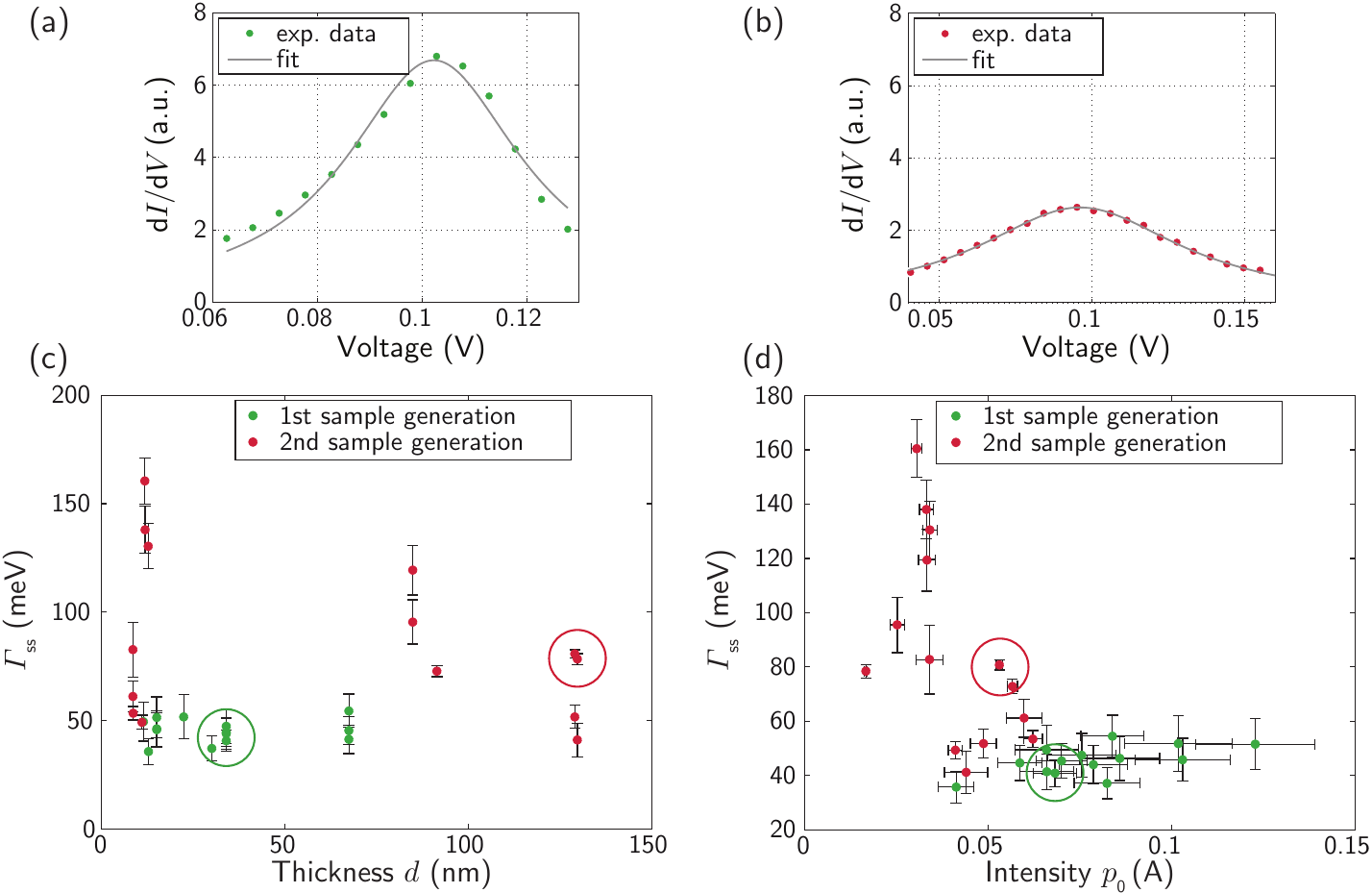}
	\caption{Characterization of the surface state for the two sample generations.
	(a,b)~Fits of a Lorentzian (Eq.~\ref{eq:didu_surf_state}) to experimental $\mathrm dI/\mathrm dV$-curves. 
	(a)~$\mathrm dI/\mathrm dV$-curve (dots) taken on a 34~nm thick La island of the 1$^\mathrm{st}$ sample generation. Fit results (straight line): \mbox{$p_0 = (0.068 \pm 0.006)$~A}, $V_0 = (102.2 \pm 1.4)$~mV, $\Gamma_\mathrm{ss} = (40.8 \pm 4.9)$~meV. 
	(b)~$\mathrm dI/\mathrm dV$-curve (dots) on a 129~nm thick island of the 2$^\mathrm{nd}$ sample generation. Fit results (straight line): $p_0 = (0.053 \pm 0.001)$~A, \mbox{$V_0 = (96.2 \pm 0.5)$~mV}, $\Gamma_\mathrm{ss} = (80.8 \pm 1.9)$~meV. (c,d)~Compilation of fit results including additional measurements. Data from 1$^\mathrm{st}$ (2$^\mathrm{nd}$) generation samples are marked by green (red) dots. The particular results shown in (a,b) are marked by colored circles.	(c)~Width (FWHM) $\Gamma_\mathrm{ss}$ vs. island thickness. (d)~$\Gamma_\mathrm{ss}$ vs. intensity $p_0$. (a-d)~Tunneling parameters: $I_\mathrm{stab} = 500$~pA at $V_\mathrm{stab} = +1$~V, \mbox{$V_\mathrm{mod}$ = 1~mV}, $T = 1.2$~K.}
	\label{fig:S1}
\end{figure}

Now, the finite lifetime of the surface state $\tau_\mathrm{ss}$ is taken into account. The lifetime broadening $\Gamma_\mathrm{ss} = \hbar/\tau_\mathrm{ss}$ consists of electron-electron ($\Gamma_\mathrm{ss}^{e-e}$), electron-phonon ($\Gamma_\mathrm{ss}^{e-\mathrm{ph}}$), and defect scattering ($\Gamma_\mathrm{ss}^\mathrm{def}$), that all add up to the overall broadening $\Gamma_\mathrm{ss} = \Gamma_\mathrm{ss}^{e-e} + \Gamma_\mathrm{ss}^{e-\mathrm{ph}} + \Gamma_\mathrm{ss}^\mathrm{def}$. The energy dependence of these different scattering channels is neglected for simplicity. Therefore, the $\delta$-function needs to be replaced by a Lorentzian \cite{Wegner_all}, which leads to
\begin{equation}
	\frac{\mathrm dI}{\mathrm dV}(V) = \int \frac{p_0\,\Gamma_\mathrm{ss}}{(E-E_0)^2+(\Gamma_\mathrm{ss}/2)^2}\, f'(E-eV)\, \mathrm d E\ .
\end{equation}
$p_0$ is a factor of proportionality, which is called intensity in the following. The width (FWHM) of the differentiated Fermi function $f'(E-eV,T)$ is about $3.5\,k_\mathrm BT = 0.36$~meV at $T = 1.2$~K, which is orders of magnitude smaller compared to the energy scale of the surface state. 
Therefore, $f'(E-eV,T)$ is replaced by a $\delta$-function $\delta(E-eV)$, resulting in
\begin{equation}
	\frac{\mathrm dI}{\mathrm dV}(V) = \frac{p_0\,\Gamma_\mathrm{ss}\, e}{(eV-eV_0)^2+(\Gamma_\mathrm{ss}/2)^2}\ .
	\label{eq:didu_surf_state}
\end{equation}
When fitting this formula to the experimental $\mathrm dI/\mathrm dV$-curves, there are three free parameters: intensity $p_0$, peak maximum $eV_0 = E_0$, and width (FWHM) $\Gamma_\mathrm{ss}$. Before fitting, the individual STS curves were normalized to have the same differential conductance at $V = +1$~V. In addition, a constant offset was subtracted, which reflects tunneling into bulk states. As depicted by the fitting result for two exemplary measured curves in Fig.~\ref{fig:S1}(a,b), this procedure leads to a good fit of the experimental data. 

The fit results for all measured $\mathrm dI/\mathrm dV$-spectra are shown in Fig~\ref{fig:S1}(c,d) color-coded according to the sample generation. $\Gamma_\mathrm{ss}$ is found to be almost independent of the island thickness. Even the thinnest films have a well-pronounced surface state, which is in agreement with previous studies \cite{Wegner_all}. 
The 1$^\mathrm{st}$ sample generation always exhibits a very narrow surface state peak with a maximum at $V_0 = +(101.7 \pm 1.2)$~mV and a width given by $\Gamma_\mathrm{ss} = (45.4 \pm 1.5)$~meV, corresponding to a lifetime of $\tau_\mathrm{ss} = \hbar/\Gamma_\mathrm{ss} = (14.5 \pm 0.5)$~fs. 
In a previous analysis of the width of the experimental surface state peak, which even considered energy-dependent lifetime broadening, a value of $\Gamma_\mathrm{ss}^\mathrm{lit} = (49 \pm 10)$~meV was reported \cite{Wegner_all}. 
In comparison with the result reported here for the 1$^\mathrm{st}$ sample generation, this implies that no additional energy broadening arises due to impurity-induced scattering. It is therefore assumed that the defect-induced broadening for this very clean sample generation can be completely neglected, and hence $\Gamma_\mathrm{ss} = \Gamma_{e-e} + \Gamma_{e-\mathrm{ph}}$ is the ``intrinsic'' lifetime broadening. 
In summary, this shows that La(0001) surfaces of the 1$^\mathrm{st}$ sample generation are of the same or even better quality compared to previous studies \cite{Wegner_all}. 

The STS spectra taken on the La islands of the 2$^\mathrm{nd}$ sample generation exhibit surface state peaks with a reduced intensity and a larger width [Fig.~\ref{fig:S1}(b-d)]. Therefore, the modified $\Gamma_\mathrm{ss}^{*} = \Gamma_{e-e} + \Gamma_{e-\mathrm{ph}} + \Gamma_\mathrm{def}^{*}$ must be affected by a defect-induced contribution $\Gamma_\mathrm{def}^{*}$. This quantity is about twice as large as the ``intrinsic'' broadening $\Gamma_\mathrm{ss}$. The influence of surface contamination on superconductivity is analyzed in the main part of the paper.


\section*{References}

\begin{thebibliography}{10}
\bibitem{Shen1972} 			
Shen L Y L 1972 \textit{AIP Conf. Proc.} \textbf{4} 31
\bibitem{Lou1972} 			
Lou L and Tomasch W 1972 \textit{Phys. Rev. Lett.} \textbf{29} 858
\bibitem{Wuehl1973} 		
W{\"u}hl H, Eichler A and Wittig J 1973 \textit{Phys. Rev. Lett.} \textbf{31} 1393
\bibitem{Levinstein1967}
Levinstein H J, Chirba V G and Kunzler J E 1967 \textit{Phys. Lett.} \textbf{24A} 362
\bibitem{deLozanna1985}
de Lozanne A L, Elrod S A and Quate C F 1985 \textit{Phys. Rev. Lett.} \textbf{54} 2433
\bibitem{Eom2006}
Eom D, Qin S, Chou M-Y and Shih C K 2006 \textit{Phys. Rev. Lett.} \textbf{96} 027005
\bibitem{Zhang2010}
Zhang R \textit{et al} 2010 \textit{Nature Phys.} \textbf{6} 104
\bibitem{Ozer2007}			
\"{O}zer M M, Jia Y, Zhang Z, Thompson J R and Weitering H H 2007 \textit{Science} \textbf{316} 1594
\bibitem{Buzea2005}
Buzea C and Robbie K 2005 \textit{Supercond. Sci. Technol.} \textbf{18} R1
\bibitem{Johnson1967}		
Johnson D and Finnemore D K 1967 \textit{Phys. Rev.} \textbf{158} 376
\bibitem{Maple1969}
Maple M B, Wittig J and Kim K S 1969 \textit{Phys. Rev. Lett.} \textbf{23} 1375
\bibitem{Pan1980} 			
Pan P H, Finnemore D K, Bevolo A J, Shanks H R, Beaudry B J, Schmidt F A and Danielson G C 1980 \textit{Phys. Rev. B} \textbf{21} 2809
\bibitem{Nixon2008}	 		
Nixon L W, Papaconstantopoulos D A and Mehl M J 2008 \textit{Phys. Rev. B} \textbf{78} 214510
\bibitem{Bagci2010} 		
Ba{\v{g}}c\i~S, T\"{u}t\"{u}nc\"{u} H M T, Duman S and Srivastava G P 2010 \textit{Phys. Rev. B} \textbf{81} 144507
\bibitem{Simonin1986} 	
Simonin J 1986 \textit{Phys. Rev. B} \textbf{33} 7830
\bibitem{SPECS}
SPECS Surface Nano Analysis GmbH, Berlin, Germany
\bibitem{Bode2007}
Bode M, Krause S, Berbil-Bautista L, Heinze S and Wiesendanger R 2007 \textit{Surf. Sci.} \textbf{601} 3308
\bibitem{MateckChempur}
ChemPur GmbH, Karlsruhe, Germany; MaTecK GmbH, J\"ulich, Germany
\bibitem{Wiebe2004}
Wiebe J, Wachowiak A, Meier F and Wiesendanger R 2004 \textit{Rev. Sci. Instrum.} \textbf{75} 4871
\bibitem{Wegner_all}
Wegner D, Bauer A, Koroteev Y M, Bihlmayer G, Chulkov E V, Echenique P M and Kaindl G 2006 \textit{Phys. Rev. B} \textbf{73} 115403; 
Wegner D, Bauer A, Kaindl G 2007 \textit{Phys. Rev. B} \textbf{76} 113410; 
Wegner D 2004 PhD Thesis Freie Universit\"at Berlin
%
\bibitem{Dynes1978}
Dynes R C, Narayanamurti V and Garno J P 1978 \textit{Phys. Rev. Lett.} \textbf{41} 1509
\bibitem{Dickey1968}		
Dickey J and Paskin A 1968 \textit{Phys. Rev. Lett.} \textbf{21} 1441
\bibitem{Brun2009}			
Brun C, Hong I-P, Patthey F, Sklyadneva I Y, Heid R, Echenique P M, Bohnen K P, Chulkov E V and Schneider W-D 2009 Phys. Rev. Lett. \textbf{102} 207002
\bibitem{Fabrega2011a}	
F{\`{a}}brega L, Cam\'{o}n A, Fern\'{a}ndez-Mart\'{\i}nez I, Ses\'{e} J, Parra-Border\'{\i}as M, Gil O, Gonz\'{a}lez-Arrabal R, Costa-Kr\"{a}mer J L and Briones F 2011 \textit{Supercond. Sci. Technol.} \textbf{24} 075014
\bibitem{Song2011a}			
Song C-L, Wang Y-L, Jiang Y-P, Li Z, Wang L, He K, Chen X, Ma X-C and Xue Q-K 2011 \textit{Phys. Rev. B} \textbf{84} 020503
\bibitem{Nepijko2000}		
Nepijko S, Getzlaff M, Pascal R, Zarnitz C, Bode M and Wiesendanger R 2000 \textit{Surf. Sci.} \textbf{466} 89
\bibitem{Getzlaff1999} 	
Getzlaff M, Bode M, Pascal R and Wiesendanger R 1999 \textit{Phys. Rev. B} \textbf{59} 8195

\end{thebibliography}
\end{document}